\documentclass[prx,superscriptaddress,floatfix,showpacs,aps,reprint,letterpaper,nobalancelastpage,footinbib]{revtex4-1}

\usepackage{amsmath}
\usepackage{amsfonts}
\usepackage{graphicx}
\usepackage[dvipsnames]{xcolor}
\usepackage[urlcolor=blue, hyperindex, colorlinks, bookmarks=true]{hyperref}

\newcommand{\op}[1]{{\hat{#1}}}
\newcommand{\sop}[1]{{\mathcal{#1}}}
\newcommand{\e}{{\mathrm{e}}}
\newcommand{\vvec}[1]{{{\mathrm{vec}(#1)}}}

\DeclareMathOperator{\tr}{tr}

\newcommand{\I}{{{\mathbb I}}}

\newcommand{\im}{{\mathrm i}}
\newcommand{\A}{{{\mathsf A}_4}}

\usepackage{microtype}

\begin{document}

\title{Demonstration of Robust Quantum Gate Tomography via Randomized Benchmarking}
\date{\today}

\author{Blake R. Johnson}
\author{Marcus P. da Silva}
\author{Colm A. Ryan}
\affiliation{Raytheon BBN Technologies, Cambridge, MA 02138, USA}
\author{Shelby Kimmel}
\altaffiliation{Current affiliation: Joint Center for Quantum Information and Computer Science (QuICS), University of Maryland, College Park, MD 20742}
\affiliation{Raytheon BBN Technologies, Cambridge, MA 02138, USA}
\affiliation{Center for Theoretical Physics, MIT, Cambridge, MA 02139, USA}
\author{Jerry M. Chow}
\affiliation{IBM T.J. Watson Research Center, Yorktown Heights, NY 10598, USA}
\author{Thomas A. Ohki}
\affiliation{Raytheon BBN Technologies, Cambridge, MA 02138, USA}

\begin{abstract}
  Typical quantum gate tomography protocols struggle with a
  self-consistency problem: the gate operation cannot be
  reconstructed without knowledge of the initial state and final
  measurement, but such knowledge cannot be obtained without
  well-characterized gates. A recently proposed technique, known as
  {\em randomized benchmarking tomography} (RBT), sidesteps this
  self-consistency problem by designing experiments to be insensitive
  to preparation and measurement imperfections. We implement this
  proposal in a superconducting qubit system, using a number of
  experimental improvements including implementing each of the
  elements of the Clifford group in single `atomic' pulses and custom
  control hardware to enable large overhead protocols. We show a
  robust reconstruction of several single-qubit quantum gates,
  including a unitary outside the Clifford group. We demonstrate that
  RBT yields physical gate reconstructions that are consistent with
  fidelities obtained by randomized benchmarking.
\end{abstract}

\maketitle

\section{Introduction}

All approaches to quantum tomography are forced to make trade-offs given the
exponentially increasing resources necessary as the size of the system
grows. There has been an aggressive effort from the community to
explore alternative approaches that return coarse-grained information
in exchange for shorter run times~\cite{ESM+07,SLP11,MSRL12}. All
these techniques rely on {\em some} assumptions about the system being
characterized. While quantum process tomography (QPT) has been shown
to suffer from systematic errors due to incorrect or unverified
assumptions about preparation and measurement~\cite{Merkel2013},
randomized benchmarking (RB) is insensitive to this ignorance and
robust against imperfections in the other operations used in the
protocol~\cite{MGE11,MGE12}. The trade-off is that randomized
benchmarking only provides information about how far away an
experiment is from an ideal Clifford group operation, i.e., the average
fidelity. In applications where a more complete reconstruction of the
operation is necessary, e.g., for debugging purposes, RB fails to
provide enough information, while the systematic errors in QPT preclude
accurate results.

Randomized benchmarking tomography (RBT)~\cite{Kimmel:2013vn} is a
recent proposal for near-complete process tomography that inherits the
robustness of standard RB and its insensitivity to state preparation
and measurement ignorance. Most notably, this technique also allows
for the estimation of the average fidelity of any applied gate relative
to \emph{any} unitary operation---in some cases, this estimation can
even be done with a polynomial number of experiments.

In this Letter, we apply RBT to reconstruct single-qubit operations in
a transmon superconducting qubit, and compare these reconstructions to
results obtained via QPT. In particular, we take advantage of the RBT
protocol to robustly reconstruct a $\pi/6$ rotation that lies outside
the Clifford group. We show that while QPT yields strong non-physical
features due to systematic errors, RBT reconstructions remain
physical. Moreover, the fidelities estimated by RBT are compatible
with fidelities estimated by standard RB.

Unsurprisingly, extracting more information requires more experiments.
Like standard QPT and other recent methods for improving upon
it~\cite{Merkel2013,BGN+13,Sta14}, RBT comes with an exponential
overhead in the total number of experiments. Occasionally, the
additional run time leads to drift in parameters of the operation or
in state-preparation and measurement errors. This may break a
fundamental assumption of most protocols that the parameters are fixed
for all rounds of the experiment. Consequently, we describe some
strategies for dealing with large experiment-count protocols,
including the use of a custom arbitrary waveform generator that
operates with very concise sequence descriptions, and readout
approaches that improve system stability.

The problem of physically valid reconstructions in tomography is more
significant in certain settings. In particular, reconstruction of
unitary operations is more sensitive to this issue because such
operations are extremal in the set of physically valid operations.
Small errors, statistical or otherwise, can easily push estimates
outside the physical bounds. To ensure we are near this challenging
limit, we endeavor to implement coherence-limited single-qubit
unitaries from the Clifford group. A new method we use to achieve this
re-introduces $Z$ control to fixed-frequency qubits, creating single-pulse,
or atomic, Clifford operations that minimize the average gate time by
avoiding multi-pulse decompositions.

\section{RBT Protocol}\label{sec:protocol}

We start with a brief description of the RBT protocol. Throughout this
discussion, we denote unitary operators in the Clifford group by $\op
C_j$, and the corresponding quantum operation (superoperator) by $\sop
C_j$. Other operations are denoted by calligraphic fonts as well,
e.g., $\sop E$. The sequential composition of two operations $\sop E$
and $\sop F$ is denoted by $\sop F \, \sop E$, meaning $\sop E$ acts
first, followed by $\sop F$. This hints at the fact that operations
can be represented as matrices and operators as vectors. This
representation is known as the Liouville (or Hilbert-Schmidt)
representation, and throughout this discussion we will use the
Liouville representation in the Pauli basis~\footnote{This combination
  of choices is referred to as the Pauli-Liouville
  representation~\cite{Kimmel:2013vn}, while other works refer to
  the resulting matrices as Pauli transfer
  matrices~\cite{GCM+12}.}. Pauli group unitaries are denoted by
$X,Y,Z$, while the identity operator is denoted $\I$.

An operation $\sop{E}$ is called unital iff $\sop E(\I)=\I$.  If $\sop
E$ is not unital, one can still refer to its {\em unital part}, $\sop
E'$, by ignoring the traceless components of $\sop
E(\I)$~\footnote{More precisely, if $\sop Q^{\perp}$ is the projector
  that takes an operator into its traceless components and $\sop E$ is
  a trace preserving operation, then $\sop E'=\sop E\,\sop
  Q^\perp+\sop Q$.}. This unital part of trace-preserving operations
can be decomposed into a linear combination of Clifford group
operations~\cite{Sco08,vDH11,Kimmel:2013vn}. In other words, $\sop E'$
can be reconstructed from estimates of overlaps
\begin{equation}
a_j = \tr\sop{C}_j^\dagger\,\sop{E},
\end{equation}
where $\sop{C}_j$ is a Clifford operation, as long as a sufficiently
large linearly independent set of Clifford operations is chosen. For a
single-qubit, instead of the full Clifford group we consider
\begin{align}\label{eq:a4-two-des}
\op C_1 &= \I &
\op C_2 &= \e^{-\im {\pi\over2}X} \notag\\
\op C_3 &= \e^{-\im {\pi\over2}Y} &
\op C_4 &= \e^{-\im {\pi\over2}Z} \notag\\
\op C_5 &= \e^{-\im {\pi\over3}{X+Y+Z\over\sqrt{3}}} &
\op C_6 &= \e^{-\im {2\pi\over3}{X+Y+Z\over\sqrt{3}}} \\
\op C_7 &= \e^{-\im {\pi\over3}{X-Y+Z\over\sqrt{3}}} &
\op C_8 &= \e^{-\im {2\pi\over3}{X-Y+Z\over\sqrt{3}}} \notag\\
\op C_9 &= \e^{-\im {\pi\over3}{X+Y-Z\over\sqrt{3}}} &
\op C_{10} &= \e^{-\im {2\pi\over3}{X+Y-Z\over\sqrt{3}}} \notag\\
\op C_{11} &= \e^{-\im {\pi\over3}{-X+Y+Z\over\sqrt{3}}} &
\op C_{12} &= \e^{-\im {2\pi\over3}{-X+Y+Z\over\sqrt{3}}}. \notag
\end{align}
which is a unitary 2-design embedded in the Clifford group~\cite{BDS+96,
DLT02, DCE+09}. We call this group $\A$, as it is isomorphic to the
alternating group of degree 4, i.e., the group of even permutations of
4 distinct labels.  The linear span of the operations in $\A$ is 10
dimensional (as is the linear span of the entire Clifford group for
single qubits), so in the experiments described here, we take the
first 10 of these operations as our linearly independent set. Given an
estimate of the \emph{overlap} vector $\vec{a} = \{a_1, \ldots,
a_k\}$, standard unconstrained least-squares inversion yields an
estimate of $\sop{E}'$ (See Appendix~\ref{app:inversion}).

\begin{figure}[htbp]
\includegraphics[width=\columnwidth]{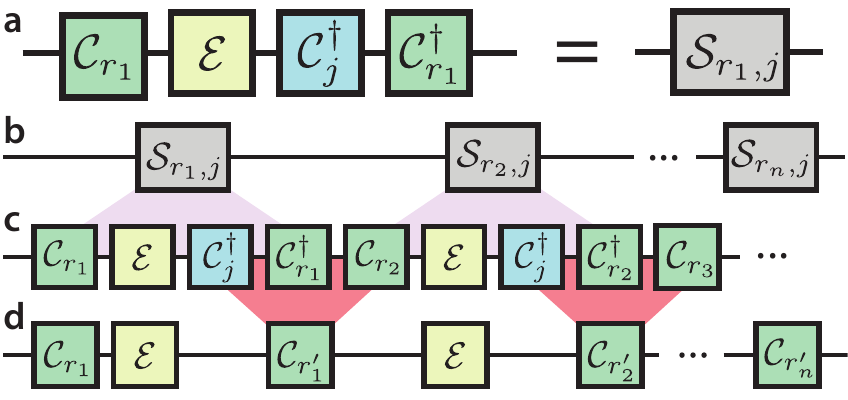}
\caption{
  \label{fig:Sequence-reduction}
  Sequence reduction of an RBT overlap experiment. \textbf{a} The
  `unit cell', $\sop{S}_{r,j}$, which is applied iteratively in an RB
  protocol, \textbf{b}. When written out \textbf{c}, one can identify
  sections with up to three Clifford operations, e.g.
  $\sop{C}_j^\dagger \, \sop{C}_{r_1}^\dagger \, \sop{C}_{r_2}$,
  which can be compiled into a single gate, $\sop{C}_{r_1'}$, also
  from the Clifford group.  The resulting sequence, \textbf{d}, has
  the same form as a standard IRB experiment, except that we allow
  $\sop{E} \not\approx \sop{C}_j$, in which case the action of the complete
  sequence may not be close to $\I$.  }
\end{figure}

We estimate the overlaps $a_j$ through interleaved RB sequences (IRB)
\cite{Magesan:2012gu,GMT+12}, as shown in
Fig.~\ref{fig:Sequence-reduction}.  That is, we iteratively apply the
sequence $\sop{S}_{r,j} = \sop{C}_r^\dagger \, \sop{C}_j^\dagger \,
\sop{E} \, \sop{C}_r$, where $\sop{C}_r$ is randomly chosen from
$\A$. We will refer to a sequence $\sop{S} = \prod_i^n
\sop{S}_{r_i,j}$ as a sequence of length $n$, where the $r_i$ are
chosen independently. In practice, it is convenient to reduce the
total sequence length by compiling the compositions of the randomly
chosen $\sop{C}_r$'s and the overlap target $\sop C_j^\dagger$, and
applying the corresponding Clifford group operation instead. Choosing
the overlap set to be a group ensures that the composed Clifford
operations are still in the same set. Consequently, the applied
sequences will take the form of alternating random Clifford
operations and the target $\sop{E}$ (see
Fig.~\ref{fig:Sequence-reduction}d). As will be discussed later, we
exhaustively sample from the set of all sequences of the form of
$\sop{S}$ for a given length, so it is advantageous to sample from a
subgroup like $\A$ instead of the full Clifford group, in order to
limit the total number of experiments.

The expectation of the fidelity between the input and the output of
this sequence, averaged over the random choices of Clifford group
operations, is~\cite{KLR+08,MGE11,MGE12}
\begin{equation}\label{eq:decay-model}
F_{j,n} = A p_j^n + B,
\end{equation}
where $p_j\in[-\frac{1}{d^2-1},1]$ is a decay rate, $d$ is the
dimension of the system (here $d=2$), and $A$ and $B$ are factors
related to preparation and measurement errors. The decay rate $p_j$ is
related to the overlap $a_j$ by
\begin{equation}\label{eq:overlap-decay}
p_j = \frac{a_j - 1}{d^2 - 1},
\end{equation}
so that estimates of the decay rates can be used to reconstruct $\sop
E$. Equivalently, the $a_j$ can be related to the average fidelity
between $\sop{E}$ and $\sop{C}_j$~\cite{HHH99,Nie02}. For small
overlaps ($a_j < 1$), the decay rates are negative, leading to
oscillatory decays in the length $n$.

Imperfections in the randomizing operations can be accounted for by
characterizing the (ideally) null operation $\sop E_0$, a zero-length
pulse~\cite{Magesan:2012gu,Kimmel:2013vn}. If only the fidelity of
$\sop E_0$ to the identity is estimated, the imperfections can only be
partially acounted for, leading to very loose bounds on the
performance of $\sop E$. However, if the unital part of $\sop E_0$ is
fully reconstructed, a much more accurate estimate can be made by
inversion~\cite{Kimmel:2013vn}. For any operation $\sop E$ that is
reconstructed via RBT, the errors can be accounted for by computing
the \emph{right} and \emph{left} corrected operations $\widetilde{\sop
  E}'_R=\sop E'\,(\sop E_0')^{-1}$ and $\widetilde{\sop E}'_L=(\sop
E_0')^{-1}\,\sop E'$, respectively. The placement of the error
operation on the left or right side of $\sop E$ is arbitrary (usually
chosen by convention), so either estimate is valid.

A difficulty in experimentally obtaining $p_j$ arises from the fact
that for most $a_j$, the resulting decay $A p_j^n$ will rapidly vanish
even for small $n$. In fact, if $\sop{E}$ is close to an ideal
Clifford operation, then the $p_j$'s in the overlap experiments
will be close to $\pm \frac{1}{3}$ or $0$, except when $\sop{C}_j
\approx \sop{E}$. For example, when $n=4$ one needs better than 1\%
precision in the measured average fidelity to distinguish
$(\frac{1}{3})^4$ from the mean value of $B$ reliably. One mitigation
is to exhaustively sample all random sequences up to a given length.
This removes configuration sampling uncertainty from the estimators of
$F_{j,n}$ above. The cost of additional experiments may be partially
offset by using control hardware with minimal overhead for uploading
gate sequences.

\section{Experimental Implementation}

We test the RBT protocol on a single qubit of a 3-transmon,
5-resonator device, described in Ref.~\cite{Chow:2012}. The probed
qubit's coherence times are $T_1 \approx 5.7\,\mathrm{\mu s}$ and
$T_2^\mathrm{echo} \approx 8.4\,\mathrm{\mu s}$, with anharmonicity
$\alpha/2\pi = -221\,\mathrm{MHz}$. The qubit's readout resonator is
coupled to a lumped Josephson parametric amplifier
\cite{Hatridge:2010ws} and pumped 17 MHz detuned from the measurement
signal to operate it in a phase-preserving mode. The readout
assignment fidelity of $\approx 95\%$ is sufficiently high that it is
advantageous to convert the measurement outcomes into binary values by
thresholding before averaging~\cite{RJG+15}. These two choices serve
to improve the system stability by reducing sensitivity to the
relative phases of the pump and measurement signals, as well as to
small voltage fluctuations in the receiver chain.

Qubit control is realized by single sideband modulation of a microwave
carrier detuned $\sim 150$MHz from the qubit transition frequency. The
shaped modulation signals are generated by a custom arbitrary waveform
generator described in Sec.~\ref{sec:hardware}. With the exception of
Z rotations which are done with a simple frame-update, we use a
fixed-duration pulse of 33.3 ns for all single-qubit gates, and vary
the control amplitudes to implement different rotations.

\subsection{Atomic Clifford Group Operations}

Quantum control in superconducting qubits is usually relaxation
limited and so to minimize gate errors it is desirable to keep the
gates as short as possible. Typical implementations consider only
control about axes in the $XY$ plane or treat $Z$ rotations separately
and are forced to decompose rotations about an arbitrary axis into a
sequence of rotations---the so-called Euler angle decomposition
\cite{Nielsen:2011:QCQ:1972505}. A relevant example of gates that
require off-axis rotations are single-qubit Clifford group operations.
These can be described as rotations about symmetry axes of the cube in
the Bloch sphere, and the cube has symmetries for $\pi$ rotations about
the $(1,0,1)$ axis and $2\pi/3$ rotations about the $(1,1,1)$ axis.
When implemented with only $XY$ control these can take up to three
times longer to implement. Here we show that arbitrary single-qubit
gates are possible with a single pulse using conventional control
schemes under mild assumptions about the linearity of the control.
We term these atomic operations.

In the reference frame rotating at the microwave control frequency,
the control Hamiltonian rotates the qubit about a fixed axis in the
$XY$-plane. If the qubit is detuned from the microwave drive, then the
total Hamiltonian picks up an additional $Z$ component and the effective
rotation axis is the vector sum of the drive and detuning terms,
giving an arbitrary effective rotation axis. This off-resonance
component can be induced by changing the frequency of the qubit or of
the microwave drive. Variable frequency qubits struggle to obtain
fine-frequency control and introduce non-Markovian effects from the
flux bias line. On the other hand, rapidly changing the frequency of a
microwave source in a phase coherent manner is a technical challenge.
However, experiments already typically provide arbitrary amplitude and
phase microwave control with an IQ mixer. This allows us to implement
a discrete-time version of a frequency change by linearly ramping the
phase of the shaped microwave drive~\cite{Patt199294}.

The effect of phase ramping on detuning is straightforward to derive
from a Trotter expansion of a tilted rotation angle gate. Consider a
Hadamard rotation ($\pi$ rotation about the $X+Z$, or $(1,0,1)$ axis).
The unitary is given by
\begin{equation}
U_\text{Had} = e^{-i\frac{\pi}{2}\frac{1}{\sqrt{2}}\left(X+Z\right)}.
\end{equation}
We can then consider a Trotter expansion of the anti-commuting $X$ and
$Z$ terms with $\theta = -i\frac{\pi}{2\sqrt{2}}$
\begin{align}
U_\text{Had} &= \lim_{n\rightarrow\infty}\left[e^{\frac{\theta}{n}Z}e^{\frac{\theta}{n}X}\right]^n \label{trotter-expansion} \\
 &= \dots e^{\frac{\theta}{n}Z}e^{\frac{\theta}{n}X}e^{\frac{\theta}{n}Z}e^{\frac{\theta}{n}X}e^{\frac{\theta}{n}Z}e^{\frac{\theta}{n}X} \notag \\
&= \dots e^{\frac{\theta}{n}Z}e^{\frac{\theta}{n}X}e^{\frac{\theta}{n}Z}\left(e^{\frac{\theta}{n}Z}e^{-\frac{\theta}{n}Z}\right)e^{\frac{\theta}{n}X}e^{\frac{\theta}{n}Z}e^{\frac{\theta}{n}X} \notag \\
&= \dots e^{\frac{\theta}{n}Z}\left(e^{\frac{2\theta}{n}Z}e^{-\frac{2\theta}{n}Z}\right)e^{\frac{\theta}{n}X}e^{\frac{2\theta}{n}Z}\left[e^{-\frac{\theta}{n}Z}e^{\frac{\theta}{n}X}e^{\frac{\theta}{n}Z}\right]e^{\frac{\theta}{n}X} \notag \\
&= \dots e^{\frac{3\theta}{n}Z}\left[e^{-\frac{2\theta}{n}Z}e^{\frac{\theta}{n}X}e^{\frac{2\theta}{n}Z}\right]\left[e^{-\frac{\theta}{n}Z}e^{\frac{\theta}{n}X}e^{\frac{\theta}{n}Z}\right]e^{\frac{\theta}{n}X} \notag \\
&= \lim_{n\rightarrow\infty} e^{\theta Z} \prod_{k=0}^{n-1}\left[e^{-\frac{k\theta}{n}Z}e^{\frac{\theta}{n}X}e^{\frac{k\theta}{n}Z}\right] \label{trotter-final},
\end{align}
where we have injected, in round parenthesis, identity $\pm Z$
rotation blocks. The same approach carries through for shaped pulses
with time varying amplitudes. The phase steps dynamically vary with
the pulse amplitude to maintain the same effective rotation axis.

\begin{figure}[htbp]
\includegraphics[width=\columnwidth]{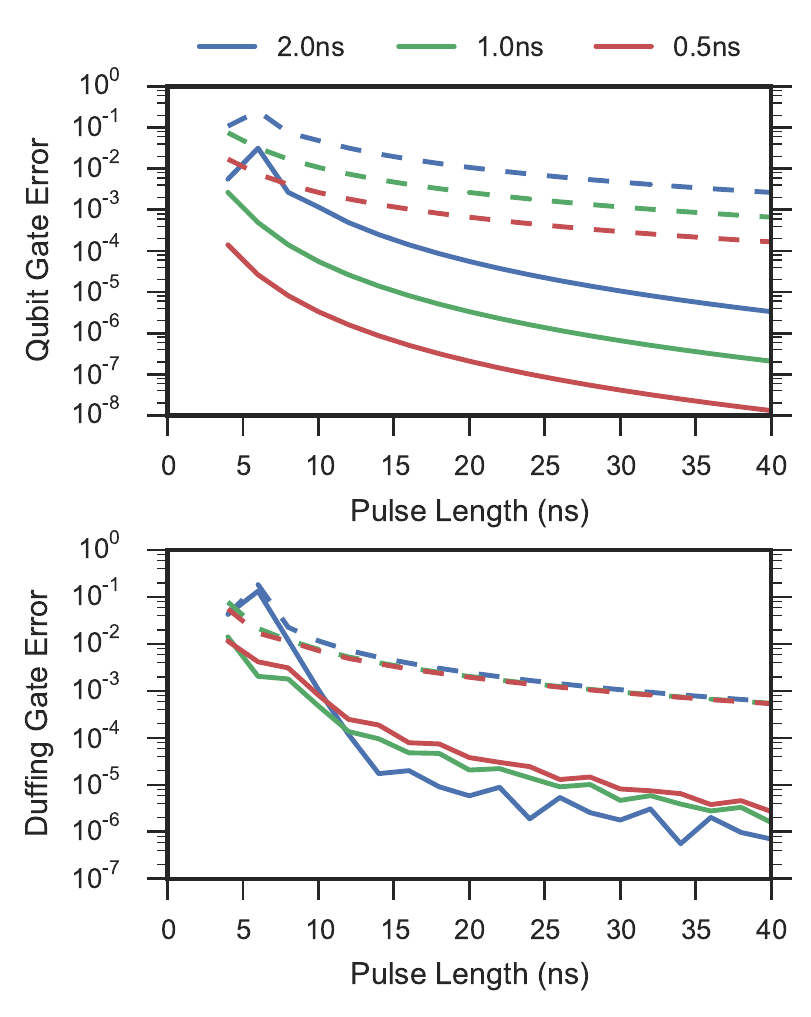}
\caption{\label{fig:AC-discretization} Simulated average gate
fidelity~\cite{Nie02} for the implementation of a Hadamard gate with a
Gaussian pulse shape extending to $\pm2\sigma$ and varying time steps
(colors). (Top) Simulates a qubit model, where the only source of
error is discretization error from implementing the frequency shift
using phase-ramps. Dashed (solid) lines indicate $1^{\text{st}}$ order
($2^{\text{nd}}$ order) Trotter approximation. (Bottom) Simulates a
5-level Duffing oscillator model of a transmon with $200\,\mathrm{MHz}$
anharmonicity. Dashed lines indicate no DRAG correction whereas solid
have $Z$-only DRAG correction. Interplay between the DRAG correction, 
phase ramping and discretization effects gives non-smooth behavior.}
\end{figure}

Truncating the product in Eq.~\eqref{trotter-final} at finite $n$,
corresponding to the number of samples in the control pulse, gives a
discrete-time implementation of a frequency shift in terms of XY
control (the X components) and per-sample frame updates (the Z
components). This is, however, only an approximation (top of
Fig.~\ref{fig:AC-discretization}). The introduced error is drastically
reduced by using a second-order Trotter expansion, or alternatively by
defining the phase of each step at the mid-point of the time bin,
rather than at the start of the ramp. This second-order version
reduces the error to a level which is insignificant compared to other
error sources in current implementations.

In implementing a detuned pulse we move to a new virtual frame where
we acquire phase at a different rate than the qubit's frame.  Thus,
when we move back to the qubit frame we must account for the accumulated
phase difference. This is represented by the final $Z$ rotation
outside the product in Eq.~\eqref{trotter-final}. Since we are already
working in a rotating frame, this $Z$ rotation may be implemented for
free by updating the phase of all subsequent pulses.

When controlling the anharmonic oscillators common to superconducting
qubit implementations, additional pulse shaping is necessary to avoid
exciting higher energy levels \cite{Gambetta2011a}. The first-order
$Z$-only correction follows through naturally to these phase ramped
pulses and their ability to demonstrate high fidelity in a Duffing
oscillator model of a transmon is shown in the bottom panel of
Fig.~\ref{fig:AC-discretization}.

\subsection{Custom Control Hardware}\label{sec:hardware}

Exhaustive sampling of even fairly short RBT overlap experiments (in
our case, up to 3 twirling gates) requires implementing thousands of
sequences of gates. This poses a practical difficulty for conventional
arbitrary waveform generators (AWGs) which require waveforms that are
the full duration of each sequence. These AWGs do not take advantage
of the relatively small number of primitives that compose RB
sequences, i.e., the pulses corresponding to each Clifford operation.
Consequently, simply uploading waveform data to a traditional AWG may
consume more wall clock time than running the experiment many
thousands of times to collect statistics.

To overcome this hurdle, we use a custom arbitrary waveform generator
called the Arbitrary Pulse Sequencer (APS). This hardware is
programmed using a natural representation for quantum information
processing experiments: it uses lists of waveform primitives (pulses
as short as 8 samples) and outputs the composite waveform produced by
concatenating these primitives without pauses or gaps between
successive waveforms. This allows the user to upload only a small set
of waveforms, such as a generating set of the Clifford group (e.g. $\I$,
$X$, $Y$, $Z$, $\sqrt{iX}$, $\sqrt{iY}$, $\sqrt{iZ}$), or the
`atomic' pulses described above, and re-use these same pulses
regardless of the sequence length. This design also has the advantage
of dramatically reducing the waveform memory requirements for the APS.
In addition, our hardware has the capability to receive new sequence
data while simultaneously outputting waveforms. We use dual-port RAM
configured as a circular buffer to fill new sequence data behind the
sequence read pointer. Consequently, data acquisition can begin with
only a small subset of the total sequence loaded onto the APS.

Fast and robust data taking is particularly important to tomography in
order to avoid unaccounted for drifts in control or sample parameters,
such as fluctuations in the qubit relaxation time. The two
improvements described above combine to significantly reduce the
overhead of experiments with large numbers of sequences and to reduce
sensitivity to drift. For example, using the APS allowed collecting an
RBT data set in $\sim 6$ hours. We estimate that with a conventional
AWG, the same experiment would take more than twice as long.

\section{Results}

\subsection{Parameter Estimation Methods}

The linear span of single-qubit Clifford group operations is 10
dimensional, and includes all trace-preserving, unital operations.
Consequently, an RBT reconstruction requires at least 10 distinct
decay experiments, where each observed decay rate $p_j$ is related to
the trace overlap $a_j$ by Eq.~\eqref{eq:overlap-decay}~\footnote{In
  principle we could reduce this to 9 distinct decays, since unital
  trace-preserving single-qubit operations have only 9 free
  parameters, but we choose not to enforce this additional constraint
  here.}. Analytical formulas relating the observed fidelities of each
sequence length to the decay rate exist~\cite{Kimmel:2013vn}, but for
the size of the statistical ensemble available to us and the
experiment signal-to-noise ratio, this procedure results in large
error bars. We remedied this by observing that the fit parameters are
not independent across all experiments. In particular, the scaling $A$
and offset $B$ for the decay curves (see Eq.~\eqref{eq:decay-model})
should be the same across different experiments as long as the
characteristics of the state preparation and measurement are stable.

The overlaps with ``instantaneous decay'' ($p_j=0$) suffer from
fitting degeneracy between $p_j=0$ and poor preparation and
measurement ($A=0$). To break the degeneracy, we simultaneously fit a
reference slow decay rate ($p_j \approx 1$) with each overlap and
require the $A$ and $B$ values to be consistent. An appropriate
reference comes from a standard RB experiment that estimates the
fidelity of the null operation $\sop E_0$, which usually has high
fidelity to the identity and therefore leads to a slow decay.

Thus, each decay rate $p_j$ is found using a four parameter fit of both
$F_{j,n}$ and another reference decay, where the parameters are: the
reference decay rate (unused in the reconstruction), the decay rate
$p_j$, a shared scale parameter $A$, and a shared offset parameter
$B$, as in Eq.~\eqref{eq:decay-model}. Moreover, because fast decays
only lead to a small number of reliable observations, while slow
decays lead to many, the figure of merit used in the joint fits is the
sum of the mean squared errors of each of the two decays.

\begin{figure}[htbp]
\includegraphics[width=\columnwidth]{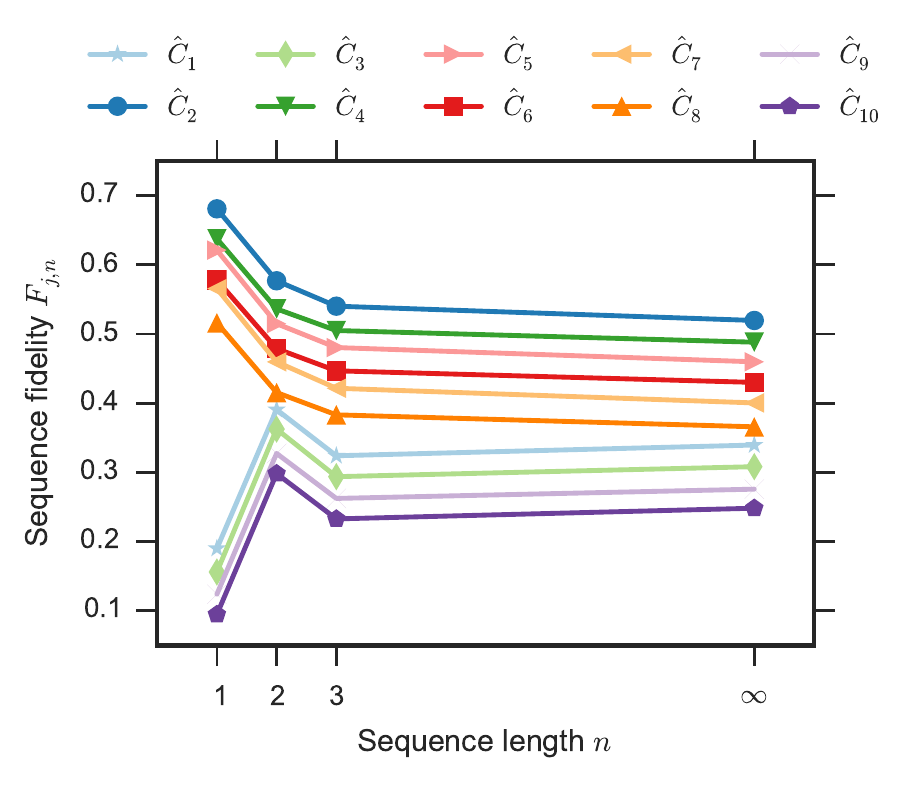}
\caption{\label{fig:decay-curves} Experimental decay curves from RBT
overlap experiments  of a Hadamard gate, vertically offset by 0.03 for
clarity. ``Infinite'' length sequences were approximated by averaging
outcomes from applying single pulses from $\A$. The curves decay
rapidly with a rate $|p_j| \sim \frac{1}{3}$, and thus the
fitting procedure requires more care than standard RB fitting
procedures, since only a few points are statistically significant.}
\end{figure}

The sequence lengths used to estimate each overlap were 1, 2, 3 and
$\infty$, where the average fidelities of infinite length sequences
were approximated by averaging the single-pulse sequences consisting
of the 12 elements of $\A$ for the same fixed initial state, which
effectively implements a twirl of the initial state. This results in a
total of $12+12^2+12^3+12=1{,}896$ different sequences (length 1 and
$\infty$ sequences were repeated 12 times for a total of $2{,}160$
experiments). The resulting decay curves for our implementation of the
Hadamard are given in Fig.~\ref{fig:decay-curves}. Since Hadamard
$\notin \A$, every $p_j < 1$, and there is no slow decay. In fact, the
decay rates $p_j \approx \pm \frac{1}{3}$. The curves with $p_j < 0$
are particularly unusual compared to standard RB experiments due to
the oscillatory behavior of the sequence fidelity with length
$n$, which occurs for vanishing overlaps $a_j \approx 0$.

For each RB, IRB, or RBT sequence we collect 10,000 repetitions,
binned into groups of 100. This binning reveals the underlying
distribution of the experimental noise, and allows one to resample the
data to create bootstrapped confidence intervals. The choice of 100
shots per bin is a trade-off between information about the
distribution versus data storage requirements and experiment runtime.

In the analysis here, the fits to the exponential decays were
performed by a non-linear least-squares (NLLS) minimization, using
Broyden-Fletcher-Goldfarb-Shanno (BFGS) minimization of the joint
figure of merit with a starting point obtained from a simple Prony
estimate of slow decays~\cite{Pro1795,PP13}. Confidence intervals were
estimated by non-parametric bootstrap percentiles~\cite{ET94}, using
2000 replications obtained from 100 samples of each of the exhaustive
experimental configurations.

\subsection{Reconstruction and fidelities}

\begin{figure}[htbp]
\includegraphics[width=\columnwidth]{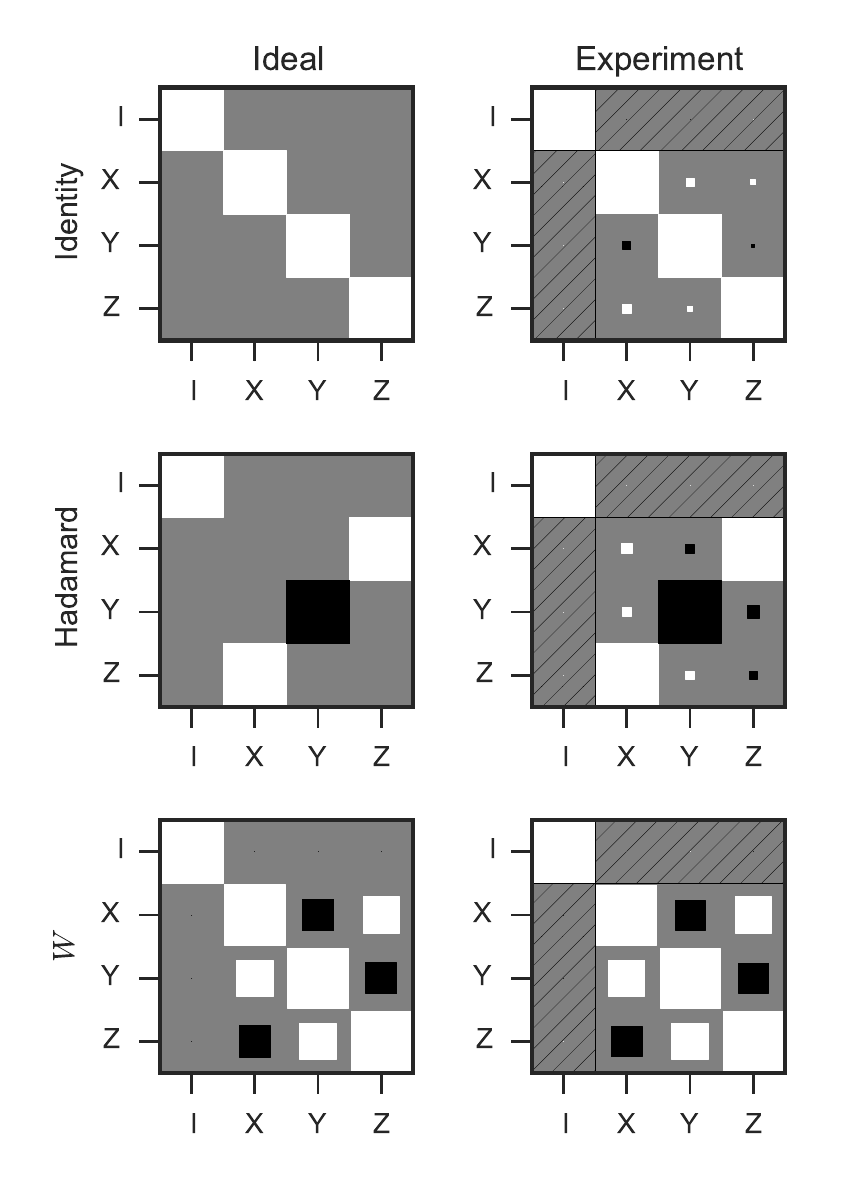}
\caption{\label{fig:reconstructions} Hinton diagrams
\cite{Hinton-1986} for the Liouville representations of the ideal
operation $\sop{E}$ (left) and reconstructed unital part of the
operation $\sop{E}'$ (right) for the identity operation, Hadamard
operation, and $\op W=\e^{-\im {\pi\over 12}{X+Y+Z\over\sqrt{3}}}$.
The area of each square corresponds to the magnitude of the
corresponding matrix element, with the sign represented by white
(positive) or black (negative). The hatched areas correspond to
parameters not accessible via the RBT protocol.}
\end{figure}

\begin{figure}[tbp]
\includegraphics[width=\columnwidth]{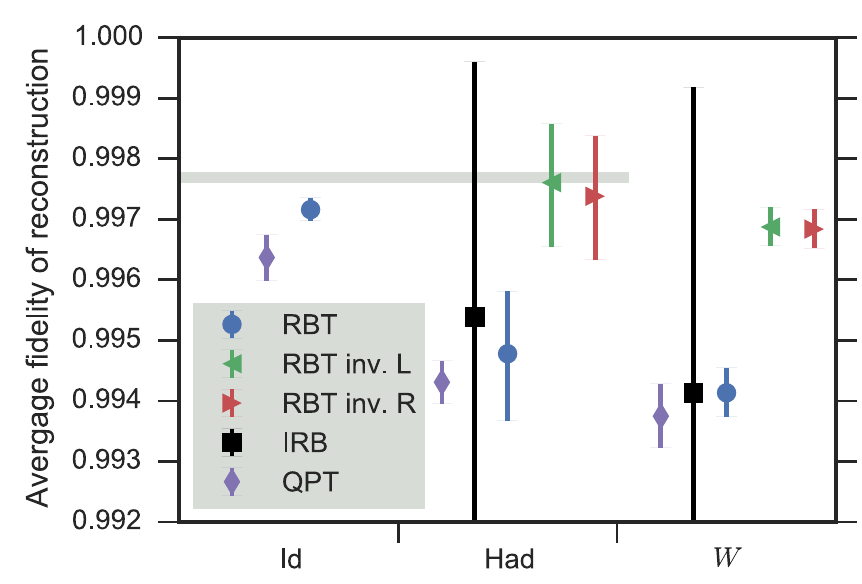}
\caption{\label{fig:fidelities} Fidelity estimates for the various
reconstructions of the identity, Hadamard, and $W$ gates. The gray bar
shows the average fidelity estimate from standard RB for the full
single-qubit Clifford group. Error bars indicate 95\% confidence
intervals for each estimate, except for the IRB points where we show
the bounds of the IRB inversion procedure. These bounds are
dramatically larger than the uncertainties in the other protocols,
extending past the bottom of the figure to roughly 0.985. After
removing randomizing error, the RBT fidelity estimate for the
Hadamard gate is consistent with standard RB. The lower fidelity of
$W$ is potentially due to the gate not being directly calibrated.
Coherence-limited control is estimated to produce gate fidelities of
0.9974.}
\end{figure}

In order to test the protocol, we apply RBT to an implementation of
the identity (a zero-length null operation), a Hadamard gate, and the
$W$ gate (a $\frac{\pi}{6}$ rotation about the $(1,1,1)$ axis, which
is a unitary operation outside the Clifford group). For each tested
gate, the fits of the overlap experiments are combined into an overlap
vector $\vec{a} = \{a_1, \ldots, a_k\}$. As described in Appendix
\ref{app:inversion}, the reconstructed operations $\sop{E}'$ are
obtained from least squares inversion. The operations, along with
their ideal, noiseless counterparts are depicted in the Pauli-Liouville
representation in Fig.~\ref{fig:reconstructions}. In this purely-real
representation the unital part excludes most elements of the first row
and column. Strictly speaking it also requires the top-left element to
be equal to 1 for a trace preserving map, although we have not
enforced this constraint in our reconstruction.

We compare RBT reconstructions to standard RB, IRB and QPT. The
average fidelities from these approaches for each of the three
operations considered are depicted in Fig.~\ref{fig:fidelities}. The
QPT results are adjusted to account for the imperfections in the
measurement, under the assumption that these imperfections are
independent of the measurement basis (i.e., the data are re-scaled so
that the $Z$ measurement spans the range $[-1, 1]$). We also compared
the RBT reconstruction to a separate IRB estimate of the fidelity of
the Hadamard gate, and to a direct estimate of the fidelity of the $W$
gate based on a subset of the decays. Although $W$ is outside the
Clifford group, it can be decomposed into a linear combination of
Clifford operations. Namely,
\begin{equation}
  \sop W = \frac{1+\sqrt{3}}{3} \sop C_1 + \frac{1}{3} \sop C_{5} +\frac{1-\sqrt{3}}{3} \sop C_{6}.
\end{equation}
Consequently, one can estimate the fidelity to $\op W$ from overlap
experiments with just $\op C_1, \op C_{5}$, and $\op C_{6}$~\cite{Kimmel:2013vn}.
However, the resulting estimate has the same bounds as the standard
IRB protocol, which leads to significantly greater uncertainty in the
estimate compared to QPT or RBT. For reference, with our gate
durations and sample coherence times, we estimate that coherence-limited
control should lead to an average gate fidelity of 0.9974.

Clearly the IRB bounds (black bars) are much looser than the fidelity
estimates using the full reconstruction, although all fidelity
estimates lie within the IRB bounds, so they are at least consistent.
The error bars for the fidelities of the QPT and RBT reconstructions
are comparable, and their estimates lie within each other's error bars
(diamond and circle). Importantly, these QPT estimates are
non-physical, while the RBT estimates do not suffer from the same
problem. However, neither of these estimates are comparable to the
fidelity estimate for the identity obtained by RB (gray bar),
indicating that the additional error is due to the average error in
the randomizing operations, and not just the error in the gate in
question.

To fully take advantage of RBT, we use the inverse of the
reconstructed null operation $\sop{E}_0'$ to remove the error of the
randomizing operations from the other reconstructions. As noted
earlier in Sec.~\ref{sec:protocol}, there is freedom to remove this
error channel by composing the inverse null operation on the left or
right side of the characterized operation, and both may be valid.
Consequently, we show both possibilities in the fidelity and
negativity estimates. With this error removed, the RBT fidelity
estimates (red and green triangles) are much closer to the RB fidelity
estimate for the identity. In other words, RBT is able to account for
the errors in the randomizing operations without the imprecision that
the IRB bounds yield.

\subsection{Systematic errors}

Imprecise knowledge about measurement and preparation imperfections is
a significant problem in quantum process tomography, because it leads
to strong systematic errors in the reconstructions of the quantum
process~\cite{Merkel2013}. While some new techniques aim at performing
full reconstruction of all experimental components in a
self-consistent manner~\cite{DLRH10,Merkel2013,BGN+13,Sta14}, techniques
such as RB, RBT, and others aim at getting around this problem by
designing experiments that are insensitive to this
ignorance~\cite{MGE11,MGE12,Magesan:2012gu,Kimmel:2013vn,Kimmel:2015}.

In order to demonstrate the reduced systematic errors in RBT compared
to QPT, we tested the reconstructed process for characteristics such
as negative eigenvalues---which, loosely speaking, correspond to
negative probabilities, and are therefore non-physical. This technique
has been used elsewhere to test for systematic errors in the analysis
of tomographic data for quantum states~\cite{MKS+13}, but it applies
equally well in the quantum process setting, thanks to the
Choi-Jamiolkoski isomorphism~\cite{Jam72,Cho75}.  This isomorphism
makes a one-to-one correspondence between a linear quantum process
$\sop E$ and the states $J(\sop E)$ resulting from applying $\sop E$
to half of a fixed maximally-entangled state. $\sop E$ is considered
to be physical if and only if it maps physical states to physical
states even when acting on only part of a state---condition known as
{\em complete positivity} (CP)~\cite{Jam72,Cho75}. This condition is
equivalent to requiring that $J(\sop E)$ be positive (i.e., that it
has only positive eigenvalues). While RBT is only able to reconstruct
the unital part of $\sop E$, positivity of a single qubit operation is
equivalent to positivity of the unital part of that same
operation~\cite{Kimmel:2013vn}, and so these tests can be applied to
the reconstructed unital operation $\sop E'$.

Following Ref.~\cite{MKS+13}, we test $\sop E'$ for non-physicality by
adaptively estimating the most-negative component of the process and
cross-validating it. For each configuration of both RBT and QPT
experiments, we divide the measurements into two halves. The first
half is used to reconstruct the unital part of the operation, which we
denote $\sop E'_1$. We compute the eigenvector of $J(\sop E'_1)$
corresponding to its most negative eigenvalue---this is what we call
{\em the negativity witness}. The second half is used to obtain an
independent estimate of the operation, which we call $\sop E'_2$, to
estimate the expectation value of the negativity witness under $J(\sop
E'_2)$. Since the parameters of $\sop E'_i$ are estimated by NLLS
instead of projective measurements on $J(\sop E'_i)$, we cannot easily
use the Hoeffding bounds in Ref.~\cite{MKS+13}. Instead, we compute
confidence regions using non-parametric bootstrap percentiles, re-sampling
only the second half of the samples while holding the negativity
witness fixed.

\begin{figure}[htbp]
\includegraphics[width=\columnwidth]{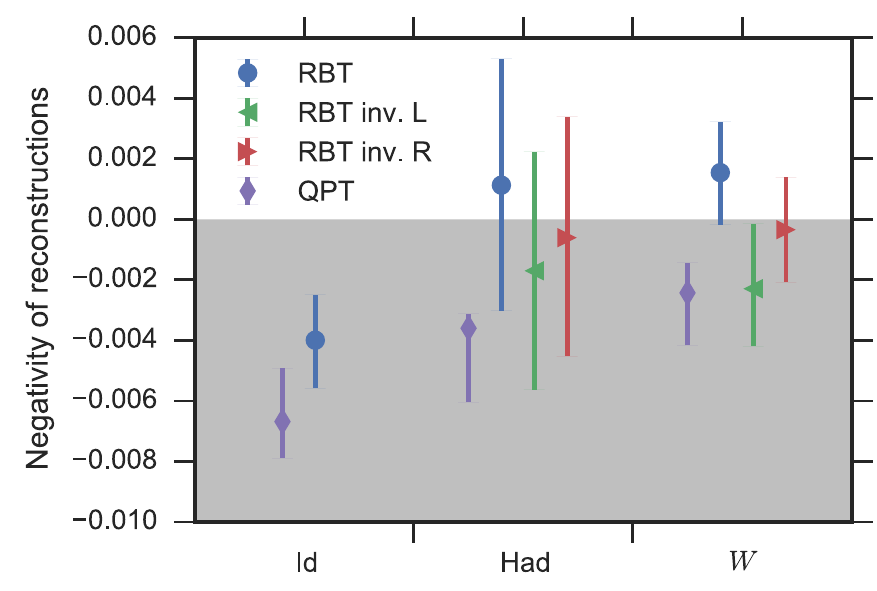}
\caption{\label{fig:negativity}Expectation value of the negativity
  witness for reconstructed operations using RBT and QPT. The error
  bars correspond to $95\%$ bootstrapped confidence intervals using
  2000 replications of the 50 samples for each experimental
  configuration used in estimating the expectation value of the
  entanglement witness.}
\end{figure}

As Fig.~\ref{fig:negativity} illustrates, the QPT reconstructions have
strongly negative eigenvalues even when statistical fluctuations are
taken in to account---the $95\%$ confidence intervals are well below
zero. With the exception of the identity, all RBT reconstructions are
consistent with CP operations---even when the non-physical estimates of the
identity are used to separate the error of the randomizing operations
from the errors in the gate itself.


The likely culprit for the observed negativity of the RBT estimates is
bias in the NLLS estimation of the decay constants. In order to test
this, we ran numerical experiments with depolarizing noise leading to
fidelities of similar magnitude to what we observed in the experiment,
as well as single-shot measurements with probability of error
comparable to our experiments. Running our estimation procedure on
this artificial data, we also found negativity for the identity
reconstruction, with similar negativity to the experiment.

It is possible to obtain physical QPT estimates from unconstrained
linear inversion by not compensating for measurement imperfections.
This leads to reconstructions without any measurable negativity, at
the cost of fidelity estimates that are in the neighborhood of $95\%$.
However, this is completely inconsistent with fidelity estimates from
RB, so they remain implausible. In other words, QPT estimates that
are consistent with RB have statistically significant unphysical
properties. By the same token, QPT estimates that are physical are
inconsistent with fidelities obtained from RB. RBT estimates, on
the other hand, are consistent with physical evolution as well as our
best estimates of fidelity for the same gates.

\section{Summary}

We have empirically demonstrated the feasibility of RBT
reconstructions of arbitrary single-qubit unitaries. These
reconstructions shows significant advantages over standard tomographic
reconstruction of quantum operations. Namely, fidelity estimates of
the RBT reconstruction are consistent with fidelity estimates obtained
through robust methods, while the reconstructed operation is
statistically consistent with a physical operation even though such a
constraint was not imposed in our reconstruction. We also demonstrated
that standard tomographical reconstructions do not satisfy these
requirements simultaneously without significant modifications (e.g.,
using gate-set tomography). However, RBT imposes large costs in terms
of experimental runtime and additional analysis complexity. Extending
this work to two-qubit process tomography would require either
daunting experiment counts for exhaustive sampling, accepting sampling
variance in the decay curves, or a modified protocol that yields slow
decays that are more ammenable to fitting procedures. Ultimately,
however, it remains unclear how to use information obtained from any
of the known tomographical protocols for fine-grained debugging of
quantum devices. Continued work is necessary to find other robust
protocols that answer targeted questions about quantum operations,
such as the recent work on robust phase estimation for pulse
calibration~\cite{Kimmel:2015}.

\begin{acknowledgments}

  The data analysis was performed using code written in
  Julia~\cite{Bezanson2014}, including the NLopt
  library~\cite{Johnson:2015}, and the figures were made with
  Seaborn~\cite{Seaborn:2014} and matplotlib~\cite{Hunter:2007}. The
  authors would like to thank George A. Keefe and Mary B. Rothwell for
  device fabrication, and Diego Rist\'e for comments on the
  manuscript. This research was funded by the Office of the Director
  of National Intelligence (ODNI), Intelligence Advanced Research
  Projects Activity (IARPA), through the Army Research Office contract
  no.~W911NF-10-1-0324. All statements of fact, opinion or conclusions
  contained herein are those of the authors and should not be
  construed as representing the official views or policies of IARPA,
  the ODNI, or the U.S. Government.

\end{acknowledgments}

\bibliography{References}

\appendix

\section{Least-squares Reconstruction of the unital part $\sop E'$\label{app:inversion}}

For any given trace preserving quantum operation $\sop E$, the unital
part $\sop E'$ is linearly related to $\vec a$, as described in the
main body of the paper, by the equation
\begin{equation}
a_j = \tr\sop{C}_j^\dagger\,\sop{E}.
\end{equation}
Since the Clifford group operations $\sop{C}_j$ are unital and trace
preserving, without loss of generality, this expression can be
replaced by
\begin{equation}\label{eq:pred-row}
a_j = \tr\sop{C}_j^\dagger\,\sop{E}'.
\end{equation}
The explicit reconstruction of $\sop E'$ from $\vec a$ can be obtained
by noting that~\eqref{eq:pred-row} implies
\begin{equation}\label{eqn:linear-problem}
\vec a = P \cdot \vvec{\sop E'}
\end{equation}
where $\vvec{\sop E'}$ is the vectorization of $\sop E'$ and $P$ is
the {\em predictor matrix} defined as
\begin{equation}
  P = \left[
  \begin{array}{c}
  \vvec{\op C_1^*\otimes\op C_1}^\dagger\\
  \vvec{\op C_2^*\otimes\op C_2}^\dagger\\
  \vdots
  \end{array}
  \right].
\end{equation}
With these definitions, we have $\sop E' = P^{I} \vec a$, where
$P^{I}$ is the Moore-Penrose pseudo-inverse of $P$ (strict inversion
is not possible as $P$ is rank deficient, thanks to Clifford
operations spanning only a 10 dimensions space, instead of the 12
dimension space of general trace preserving operations, or the 16
dimensional space of general operations). In the presence of
homoscedastic statistical fluctuations, the use of $P^{I}$ corresponds
to a least-squares estimate with minimum Euclidean norm, although
solving \eqref{eqn:linear-problem} through other equivalent means is
preferable to pseudo-inversion, for reasons of numerical stability
(the backslash operator in MATLAB and Julia~\cite{Bezanson2014}, as well as specialized
functions in other software packages, provide this functionality).





\end{document}